\documentstyle[twocolumn,prl,aps,epsfig,floats,hyperref]{revtex}

\newcommand{\NPB}[3]{{\it Nucl.~Phys.}~{\bf B#1}, #2~(19#3)}
\newcommand{\PLB}[3]{{\it Phys.~Lett.}~{\bf B#1}, #2~(19#3)}

\def\D0{D\O~}

\def\beq{\begin{equation}}
\def\eeq{\end{equation}}
\def\bdm{\begin{displaymath}}
\def\edm{\end{displaymath}}
\def\bea{\begin{eqnarray}}
\def\eea{\end{eqnarray}}

\def\url#1{\mbox{\href{#1}{\sf #1}}}
\def\urll#1#2{\mbox{\href{#1}{\sf #2}}}

\begin{document}
\draft

\twocolumn[\hsize\textwidth\columnwidth\hsize\csname
@twocolumnfalse\endcsname

\title{
Diphoton Background to Higgs Boson Production at the LHC \\
with Soft Gluon Effects
}

\author{
C.~Bal\'azs, ~~ 
P.~Nadolsky, ~~ 
C.~Schmidt, ~~ and ~~ 
C.--P. Yuan} 

\address{
Michigan State University, East Lansing, MI 48824, U.S.A.}

\date{April 24, 2000}

\maketitle
\thispagestyle{empty}

\begin{abstract}

The detection and the measurement of the production cross section of a 
light Higgs boson at the CERN Large Hadron Collider demand the accurate 
prediction of the background distributions of photon pairs. To improve 
this theoretical prediction, we present the soft-gluon resummed 
calculation of the $p p \to \gamma \gamma X$ cross section, including the 
exact one loop $g g \to \gamma \gamma g$ contribution. By incorporating 
the known fixed order results and the leading terms in the higher order 
corrections, the resummed cross section provides a reliable prediction for 
the inclusive diphoton invariant mass and transverse momentum 
distributions. Given our results, we propose the search for the Higgs 
boson in the inclusive diphoton mode with a cut on the transverse momentum 
of the photon pair, without the requirement of an additional tagged jet.

\pacs{PACS number(s): 
12.38.Cy, 
13.85.Qk. 
\hfill  hep-ph/9905551, CTEQ--905, MSUHEP--90526} 
\end{abstract}

\vskip1pc]

\setcounter{footnote}{0}
\renewcommand{\thefootnote}{\arabic{footnote}}


\section{Introduction}

The direct search for the Standard Model
(SM) Higgs boson at the CERN LEP collider constrains its
mass $m_H$ to be greater than $\sqrt{S} - m_Z \sim 108$
GeV \cite{MoriondTalks}.
Meanwhile, recent electroweak global fits \cite{EWGlobalFits}, and the 
measurements of the $W^\pm$ boson and top quark masses \cite{mWmt} suggest 
that the SM Higgs mass is lower than about 260 GeV. 
If $m_H$ is less than twice
the $Z^0$ boson mass, as the electroweak fits hint, then at the CERN
Large Hadron Collider (LHC) the most promising detection modes of the SM
Higgs boson will be $p p \to H X \to \gamma \gamma X$ \cite{AtlasAndCMS} 
and the associated production
$p p \to H\, {\rm jet}\, X \to \gamma \gamma \, {\rm jet} \, X$ 
\cite{Abdullin}. According to
earlier studies, a statistical significance on the order of 5-10 can
be reached for both of these signals, actual values depending on
luminosity and background estimates. In Ref.~\cite{Abdullin} it was
also found that in order to optimize the significance it is necessary to
impose a 30 GeV cut on the transverse momentum of the jet, or
equivalently (at NLO precision), on the $Q_T$ of the photon pair. With
this cut in place extraction of the signal in the Higgs plus jet mode 
requires the precise knowledge of both the signal and
background distributions in the mid- to high-$Q_T$ region.

The precise determination of signal and background rates of the inclusive
diphoton process demands the calculation of the large QCD corrections.
Higher order corrections, both to the signal \cite{HiggsFixO} and the
background \cite{DiphotonFixO}, increase the rate significantly, by 
a factor of about 2.
In the case of the background this large
increase is mostly due to the fact that the diphoton cross section
receives a large contribution from the formally higher order
$g g \to \gamma \gamma$ partonic subprocess, which is of the same order of
magnitude as the 
$q {\bar q} + q g \to \gamma \gamma X$ subprocesses \cite{BalazsYuanZZ}. 
Since the
lowest order $g g \to \gamma \gamma$ subprocess proceeds through a box
diagram, the calculation of yet higher order corrections to this process
is complicated.

A reliable calculation of the transverse momentum distribution of the
Higgs boson or the background
photon pair also requires the resummation of the potentially large
logarithmic contributions of the type $\ln (Q/Q_T)$ (where $Q$ is the
invariant mass of the pair), arising as a result of multiple
soft-gluon emission. Using the soft-gluon resummation technique, the low-
to mid-$Q_T$ region can be predicted, and the resummed calculation can
be matched to the fixed order, giving a reliable prediction in the whole
$Q_T$ range \cite{BalazsYuanZZ,BalazsYuanH}. 
As an added bonus the resummed cross 
section also exhibits reduced scale dependence since it includes the most
important higher order contributions. It also gives a hint of the 
size of the yet higher order corrections.

The effects of the multiple soft-gluon radiation on the transverse 
momentum distribution of Higgs bosons were discussed in a recent paper 
\cite{BalazsYuanH}. In the present work, we analyze the same effects on 
the transverse momentum of the background photon pair, extending results 
previously published in Refs. \cite{BalazsYuanZZ,BalazsBergerMrennaYuan}.
When $Q_T$ is integrated, the resummed calculation of
Ref.\cite{BalazsYuanZZ} gives the rates for the $q {\bar q} + q g \to
\gamma \gamma X$ subprocesses at the ${\cal O}(\alpha^2 \alpha_S)$ precision.
That calculation also
includes the photon fragmentation contribution, and approximates the 
${\cal O}(\alpha^2 \alpha_S^3)$ $g g + q g \to \gamma \gamma X$ rate. 
In this work,
we include the exact one loop ${\cal O}(\alpha^2 \alpha_S^3)$ $g g
\to \gamma \gamma g$ calculation to improve the diphoton background
prediction in the high $Q_T$ region. The above fixed
order and resummed contributions are implemented in the ResBos
\cite{BalazsYuanWZ} Monte Carlo event generator which was used to produce our
numerical results.

\section{Analytical Results}

The one loop $gg \to \gamma \gamma g$ amplitude 
can easily be derived from the one loop five-gluon ($5g$) amplitudes
of Ref.~\cite{pentagon} for the case when the particles in the
loop are fermions in the fundamental representation, by replacing
two of the gluon vertices with photon vertices. Since the $5g$
amplitude is explicitly given in a color-decomposed form, it is possible
to replace the SU(3) generators and strong couplings ($g_S$) of two
of the gluons with U(1) generators and the electromagnetic couplings ($e$) of
photons. 
The final expression for the square of the three-gluon--two-photon
($3g2\gamma$) amplitude is 
\begin{eqnarray}
&& |A (g_1g_2 \to g_3\gamma_4 \gamma_5 )  |^2  =
8 (e Q_i)^4 g_S^6 N_C (N_C^2 -1 ) 
\times \nonumber \\ && ~~~ 
\sum_{helicities} \Biggl|\,  
\sum_{\sigma \in {\rm COP}^{(123)}_4} A_{5,1}^{[1/2]}(\sigma_1,\sigma_2,
\sigma_3,\sigma_4,\sigma_5) \Biggr|^2,
\label{EqDefA2}
\end{eqnarray}
where $\sigma_i$ is shorthand for the 4-momenta and helicities, 
$\{p_{\sigma_i},\lambda_{\sigma_i}\}$, of the gluons 1,2,3 and the photons 4,5.
The charge of the quarks in the loop is given by $Q_i$ in
the units of the charge of the positron, and $N_C=3$ is the number of
colors in QCD.
In Eq.(\ref{EqDefA2}) ${\rm COP}^{(123)}_4$ denotes the subset of 
permutations of (1,2,3,4) that leave the ordering of (1,2,3) unchanged up to
a cyclic permutation 
\begin{eqnarray}
&& \sum_{\sigma \in {\rm COP}^{(123)}_4} A_{5;1}^{[1/2]}(\sigma_1,\sigma_2,
\sigma_3,\sigma_4,\sigma_5) =                \nonumber \\ && ~~~ 
      A_{5;1}^{[1/2]}( 1, 2, 3, 4, 5)+
      A_{5;1}^{[1/2]}( 1, 2, 4, 3, 5)+ \nonumber \\ && ~~~ 
      A_{5;1}^{[1/2]}( 1, 4, 2, 3, 5)+
      A_{5;1}^{[1/2]}( 4, 1, 2, 3, 5)+ \nonumber \\ && ~~~ 
      A_{5;1}^{[1/2]}( 3, 1, 2, 4, 5)+
      A_{5;1}^{[1/2]}( 3, 1, 4, 2, 5)+ \nonumber \\ && ~~~ 
      A_{5;1}^{[1/2]}( 3, 4, 1, 2, 5)+
      A_{5;1}^{[1/2]}( 4, 3, 1, 2, 5)+ \nonumber \\ && ~~~ 
      A_{5;1}^{[1/2]}( 2, 3, 1, 4, 5)+
      A_{5;1}^{[1/2]}( 2, 3, 4, 1, 5)+ \nonumber \\ && ~~~ 
      A_{5;1}^{[1/2]}( 2, 4, 3, 1, 5)+
      A_{5;1}^{[1/2]}( 4, 2, 3, 1, 5).
\end{eqnarray}
The partial amplitudes $A_{5;1}^{[1/2]} (1,2,3,4,5)$ 
for the various helicities of the external gluons and photons
are given in Ref.~\cite{pentagon}.

The extension of the Collins-Soper-Sterman soft-gluon resummation
formalism \cite{CSS} to diphoton production was discussed in
Ref.~\cite{BalazsBergerMrennaYuan}. We follow that work when calculating
the ${\cal O}(\alpha_S)$ fixed-order corrections to the $q {\bar q} \to
\gamma \gamma$ subprocess, including the fragmentation contributions,
and resumming the contributions of the $q {\bar q} \to \gamma \gamma g$,
$q g \to \gamma \gamma q$ and ${\bar q} g \to \gamma \gamma {\bar q}$
subprocesses. 
In addition, Ref.~\cite{BalazsBergerMrennaYuan} used an approximate
expression for the cross-section of the $g g \to \gamma \gamma g$ subprocess,
that accounted for the emission of soft and/or collinear gluons off the
initial gluons, and was valid in the limit of small non-zero $Q_T$. 

Using the exact formula (\ref{EqDefA2}) for the matrix element of the
$g g \to \gamma \gamma g$ subprocess, we can now also analyze the 
high-$Q_T$ region. 
We match the above fixed-order result with the resummed results of
Ref.~\cite{BalazsBergerMrennaYuan}, following the procedure
of Ref.~\cite{BalazsYuanWZ},  
to obtain a prediction in the whole
range of $Q_T$.  
In the resummation of the $g g + q g \to \gamma \gamma g$ process, we
use the $A^{(1)}$ and $A^{(2)}$ coefficients of Ref.~\cite{CPHiggs},
since these coefficients depend only on the initial partons and are
independent of the partonic process itself. The $B^{(1)}$ coefficient
for the $g g \to \gamma \gamma g$ process is the same as the one for $g
g \to H g$, also given in Ref.~\cite{CPHiggs}. The only unknown part of the
resummed cross-section for this subprocess is the
coefficient $C^{(1)}$, which would require the computation of
two-loop virtual diagrams. Following Ref.~\cite{BalazsBergerMrennaYuan},
we approximate $C^{(1)}$ in the process $g g + q g \to \gamma \gamma g$ by 
the expression for $C^{(1)}$ in the process $g g \to H g$ in the limit of
a large top quark mass.  This approximation is reasonable when both photons
have large transverse momenta, because then the real and virtual corrections
are dominated by soft and collinear radiation off external lines, which
is identical to that in the Higgs production process. 


\begin{figure}[t]
\vspace*{-7mm}
\epsfig{file=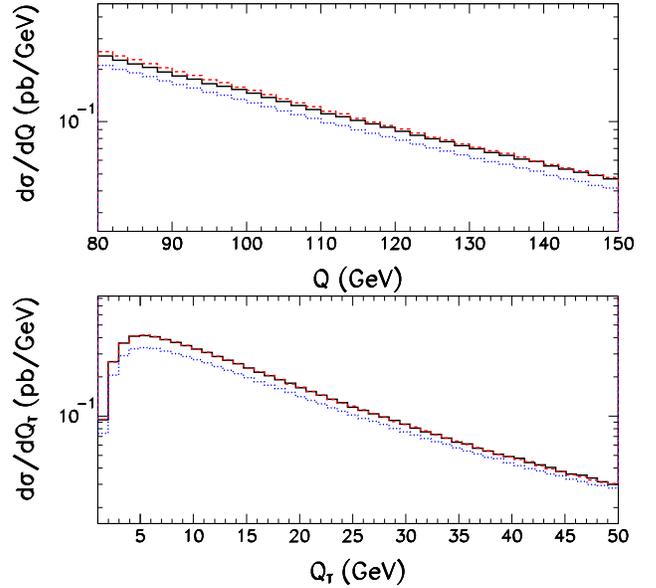,width=8.65cm,height=8.65cm}
\vspace*{1mm}
\caption[Fig:DYCFn]{

The resummed part of invariant mass and transverse momentum distributions 
of photon pairs at the LHC, calculated for the $q{\bar q} + q g \to \gamma 
\gamma X$ subprocess using ResBos. The solid curves are calculated using 
the exact Wilson coefficient $C^{(1)}$. For the dashed curves, $C^{(1)}$ 
is set identical to that of the Drell-Yan process. For the dotted curves, 
$C^{(1)}$ is set to zero.

}
\label{Fig:DYCFn}
\end{figure}

To support this statement, we examine the behavior of the analogous 
approximation for the case of the 
$q {\bar q}, q g, {\bar q} g \to \gamma \gamma X$ process. 
We calculate the resummed part of this cross section in three different 
ways: with an exact $C^{(1)}$, with an approximate $C^{(1)}$,
and without the $C^{(1)}$ coefficient. The approximate 
$C^{(1)}$ coefficient is identical to that of the Drell-Yan process, and
omits terms coming from the additional virtual corrections 
to the hard process (c.f. Eqs. (8) and (11) of 
\cite{BalazsBergerMrennaYuan}). The results are displayed in Fig. 
\ref{Fig:DYCFn}. The figure shows that the approximation (dashed curve) is
better toward the high invariant mass region, which 
is expected, since $Q$ is correlated with the transverse momentum of the 
individual photon in the central rapidity region. 
According to the invariance mass plot, the approximation
slightly overestimates the rate for low $Q$'s (by a few percent), 
and the (dotted) curve without $C^{(1)}$ 
deviates from the exact (solid) curve by about 20 percent, which is 
the size of the NLO corrections to this process. In conclusion, the 
approximate $C^{(1)}$ coefficient (borrowed from the Drell-Yan
process) captures most of the NLO corrections for 
the $q {\bar q}, q g, {\bar q} g \to \gamma \gamma X$ process,
and is a good ansatz for both
$Q$ and $Q_T$, when compared with the exact calculation.
In the absence of the exact calculation for the $C^{(1)}$ coefficient, which
requires the knowledge of the exact two loop virtual corrections to 
the $g g + q g \to \gamma \gamma$ process, 
we propose to use the approximate $C^{(1)}$ coefficient, borrowed from 
the $g g \to H X$ process, to estimate the effect of the higher
order corrections to the distributions of the photon pairs.
We expect that this should give a better estimate of the event rates than using
only the $C^{(0)}$ coefficient in our calculation.


Finally, the small $q g$ component is still 
approximated as in Ref.~\cite{BalazsBergerMrennaYuan}, noting that it
is highly suppressed
due to the large difference between the gluon and quark luminosities
in the probed region of momentum fraction. 

\section{Numerical Results}

The above described analytic results are implemented in the ResBos Monte
Carlo event generator \cite{BalazsYuanWZ}. In the numerical calculations
we use the center-of-mass energy 14 TeV and 
the following electroweak parameters \cite{PDB}: 
\begin{eqnarray}
G_F =& 1.16639\times 10^{-5}~{\rm GeV}^{-2},~~m_Z &= 91.187~{\rm GeV}, 
\nonumber \\
m_W =& 80.41~{\rm GeV}, ~~~~~~~~~~~~~~ \alpha (m_Z) &= \frac 1{128.88}.
\nonumber
\end{eqnarray}
We use the NLO expressions for the running electromagnetic and strong
couplings $\alpha(\mu)$ and $\alpha_S(\mu)$, as well as the NLO 
parton distribution function set CTEQ4M.
We set the renormalization scale equal to the
factorization scale: $\mu_R=\mu_F=Q$.
For the choice of the non-perturbative function of the resummation for the
various subprocesses, we follow Ref.~\cite{BalazsBergerMrennaYuan}.

Our kinematic cuts, imposed on the final state photons, reflect the
optimal detection capabilities of the ATLAS detector
\cite{AtlasAndCMS}:
\begin{itemize}
\item $p_T^\gamma >25$ GeV, for the transverse momentum of each photon,
\item $\left| y^\gamma \right| <2.5$, for the rapidity of each photon, 
and
\item 
$p_T^1/(p_T^1+p_T^2)<0.7$, to suppress the fragmentation contribution,
where $p_T^1$ is the transverse momentum of the photon 
with the higher $p_T$ value.
\end{itemize}
Additionally, we restrict the invariant mass of the photon pair in the
relevant region for the light Higgs production: 80 GeV $< Q <$ 150 GeV.
We also apply a $\Delta R=0.4$ separation cut on the photons, but our results
are insensitive to this cut (cf. Ref.~\cite{BalazsBergerMrennaYuan}).

\begin{figure}[t]
\vspace*{-7mm}
\epsfig{file=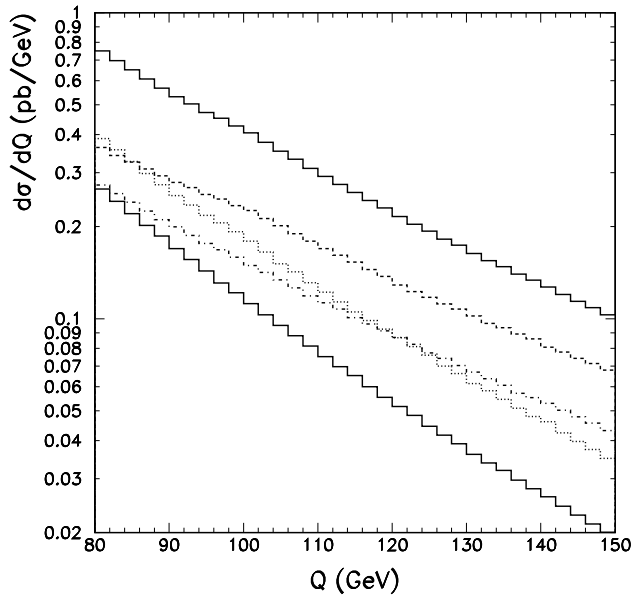,width=8.65cm,height=8.65cm}
\vspace*{1mm}
\caption[Fig:Q]{

Invariant mass distributions of photon pairs at the LHC, calculated using
ResBos. The total curve
(upper solid) is the sum of the ${\cal O}(\alpha_S)$ $q{\bar q} + q g
\to \gamma \gamma X$ (dashed) and the ${\cal O}(\alpha_S^3)$ $g g
+ q g \to \gamma \gamma X$ (dotted) contributions. The leading order curves 
for the contributions from 
${\cal O}(\alpha_S^0)$ $q {\bar q} \to \gamma \gamma $ (dash-dotted) and
${\cal O}(\alpha_S^2) $ $g g \to \gamma \gamma $
(lower solid) are also shown for comparison. 

}
\label{Fig:Q}
\end{figure}

In Fig.~\ref{Fig:Q} we display the invariant mass distribution of photon
pairs at the LHC in the inclusive process $pp \to \gamma\gamma X$, calculated
using ResBos with the above cuts. 
We present separately the resummed contribution from the subprocesses
${\cal O}(\alpha_S)$ $q{\bar q} + q g \to \gamma
\gamma X$ (dashed) which includes the fragmentation, and  
${\cal O}(\alpha_S^3)$ $g g + q g \to \gamma\gamma X$ (dotted),
as well as the total distribution (upper solid), which is the sum of these 
two curves.  We also display, for comparison, the leading order contributions
from the subprocesses ${\cal O}(\alpha_S^0)$ $q{\bar q} \to \gamma
\gamma$ (dash-dotted) and ${\cal O}(\alpha_S^2)$ $g g \to \gamma
\gamma$ (lower solid).  

The normalization of the resummed cross-section is determined order-by-order 
by the coefficients $C^{(n)}$, with the coefficient $C^{(1)}$ derived
from the NLO corrections. In Fig.~\ref{Fig:Q} we used the exact 
$C^{(1)}$ for the subprocess $q{\bar q} + q g \to \gamma \gamma X$, for which
the complete NLO cross-section is known. For the subprocess 
$g g + q g \to \gamma\gamma X$, in which the NLO virtual 
corrections have not yet been calculated, we used the approximate $C^{(1)}$ 
as described in the previous Section. 
We found that the ${\cal O}(\alpha_S^3)$ cross-section 
is sensitive to  $C^{(1)}$. Namely,
the resummed cross-section with the approximate  $C^{(1)}$ included 
is about 1.9 times larger than the one with $C^{(1)} = 0$.

In the $Q > 85$ GeV region the contribution of the
$q{\bar q} + q g \to \gamma \gamma X$ subprocess is higher than that of the
$g g + q g \to \gamma \gamma X$. From Fig.~\ref{Fig:Q} we can also read
the $K$-factors, which are defined as the ratios of the resummed to the 
leading order results. The $K$-factor for the $q{\bar q} + q g \to
\gamma \gamma X$ process (the ratio of the dashed and dash-dotted curves)
is between 1.40 and 1.70 in the invariant mass range of interest.
Similarly, the $g g + q g \to \gamma \gamma X$ $K$-factor (the ratio of
the dotted and lower solid curves) is between 1.45 and 1.75. This
results in an overall $K$-factor of 1.4 to 1.7.
These $K$-factors are about the same as the NLO/LO $K$-factors in the fixed
order perturbative calculations \cite{HiggsFixO,DiphotonFixO}.

\begin{figure}[t]
\vspace*{-7mm}
\epsfig{file=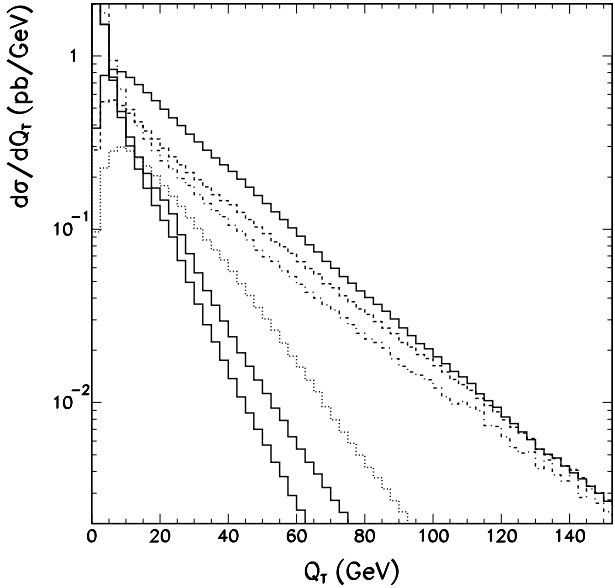,width=8.65cm,height=8.65cm}
\vspace*{1mm}
\caption[Fig:QT]{

Transverse momentum distributions of photon pairs at the LHC. The total
resummed curve (upper solid) is the sum of the resummed $q{\bar q} + q g
\to \gamma \gamma X$ (dashed), and the resummed $g g + q g \to \gamma \gamma
X$ (dotted) contributions. The fixed-order curves for the contributions from 
${\cal O}(\alpha_S)$ $q{\bar q} + q g \to \gamma \gamma X$ (dash-dotted) and 
${\cal O}(\alpha_S^3)$ $g g + q g \to \gamma \gamma X$ (middle solid) are also 
shown for comparison. The approximation for the latter used in 
\cite{BalazsBergerMrennaYuan} is also shown (lower solid).

}
\label{Fig:QT}
\end{figure}

In Fig.~\ref{Fig:QT} we plot the transverse momentum
distribution of photon pairs at the LHC.  In addition to the total
resummed result, we give the resummed and fixed-order
calculations separately for both the 
${\cal O}(\alpha_S)$ $q{\bar q} + q g \to \gamma \gamma X$ and the $
{\cal O}(\alpha_S^3)$ $gg + q g \to \gamma \gamma X$ subprocesses. 
In both
cases the resummed results deviate substantially from the fixed order 
predictions in
the $0 < Q_T < 100$ GeV region. At $Q_T = 30$ GeV the resummed curves
are higher by about 30 and 50 percent for the $q{\bar q} + q g \to
\gamma \gamma X$ and $g g + q g \to \gamma \gamma X$ subprocesses,
respectively. As a result the total resummed curve exceeds the
total fixed-order prediction by almost 40 percent at $Q_T = 30$ GeV. 
This is the $Q_T$ region where the kinematic cuts are applied in order
to optimize the statistical significance of the signal in the Higgs plus jet
mode.  
Thus, the use of the resummed prediction is necessary to extract a reliable 
statistical significance, and also to make a correct determination of the 
Higgs production cross section, in the presence of kinematic cuts.

Fig.~\ref{Fig:QT} also shows that if the photon pair is constrained to
be in the mid- to high-$Q_T$ region the contribution of the $g g + q g \to
\gamma \gamma X$ subprocess is small. At $Q_T = 40$ GeV, for example,
the $g g$ initial state accounts for less than 30 percent of the total cross
section.  In the ResBos program,
the $g g + q g \to \gamma \gamma X$ rate
is predicted purely by the resummed calculation 
and does not cross over into the fixed-order ${\cal O}(\alpha_S^3)$
calculation until after about $Q_T=100$ GeV, 
at which point this rate is negligible. 
For reference we also show the approximate fixed order curve used in 
\cite{BalazsBergerMrennaYuan}. The curve calculated using the exact matrix 
element significantly exceeds the approximation in the high $Q_T$ region.

\begin{figure}[t]
\vspace*{-7mm}
\epsfig{file=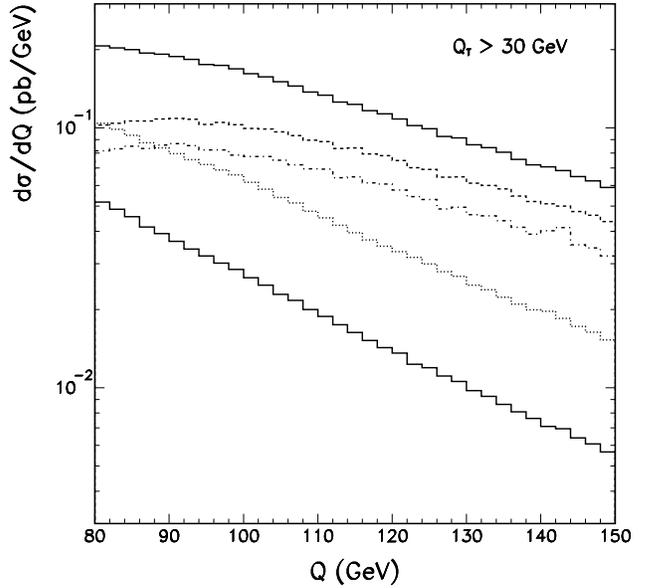,width=8.65cm,height=8.65cm}
\vspace*{1mm}
\caption[Fig:QWQTCUT]{

Invariant mass distributions of photon pairs at the LHC, 
with the cut $Q_T>30$ GeV.  The total
resummed curve (upper solid) is the sum of the resummed $q{\bar q} + q g
\to \gamma \gamma X$ (dashed), and the resummed $g g + q g \to \gamma \gamma
X$ (dotted) contributions. The fixed-order curves for the contributions from 
${\cal O}(\alpha_S)$ $q{\bar q} + q g \to \gamma \gamma X$ (dash-dotted) and 
${\cal O}(\alpha_S^3)$ $g g + q g \to \gamma \gamma X$ (lower
solid) are also shown for comparison. 

}
\label{Fig:QWQTCUT}
\end{figure}

To illustrate the higher order effects on the invariant mass
distribution in the presence of a $Q_T$ cut, in Fig.~\ref{Fig:QWQTCUT}
we plot $Q$ of the photon pair while restricting $Q_T > 30$ GeV. Due
to the different shape of the $Q_T$ distributions the cut offsets the
fixed-order and resummed rates, as explained in Ref.~\cite{BalazsYuanWZ}.
The effect is larger for the $g g$ channel, since there the resummed and
fixed-order $Q_T$ distributions deviate more, signaling higher corrections 
to the $g g$ process than to the $q {\bar q}$ process. 
In the presence of kinematic cuts the difference
in the $Q$ distribution, between the fixed order and the resummed
calculations, can be as high as 50 percent.
The $Q_T > 30$ GeV cut also suppresses the $g g$ channel, decreasing the
uncertainty of the total prediction.

\section{Conclusions}

In this paper we present the resummed calculation of the $p p \to \gamma 
\gamma X$ distributions including the exact fixed order $g g \to \gamma 
\gamma g$ contribution. Combining the known fixed order QCD corrections 
and the most important logarithmic terms of the higher order corrections, 
the resummed cross section provides a reliable prediction for the 
inclusive diphoton invariant mass and transverse momentum ($Q_T$) 
distributions. With a $Q_T$ cut the least reliable $g g \to \gamma \gamma 
X$ component can be suppressed, and the prediction further improved.

Given our results, we propose the search for the Higgs boson in the
inclusive diphoton mode with a cut on the transverse momentum of the
photon pair. 
This measurement can be done without the requirement of a tagged jet,
which is necessary in the $\gamma \gamma$ jet mode.
Therefore, it is independent of the jet algorithm used, it can be 
performed more precisely experimentally, and it can be predicted more
reliably from a resummed calculation such as presented here.

\vspace{0.4cm} 

While finishing this paper we became aware of a similar work, in which the
authors extract the $3g2\gamma$ amplitude from the $5g$ amplitude
\cite{kunszt}. Our fixed order analytical and numerical results agree
with the results of that paper.

\section*{Acknowledgments}

We thank the CTEQ Collaboration, L. Dixon and S. Mrenna for invaluable
discussions, J. Huston and M. Abolins for help with the ATLAS parameters
and for useful conversations. This work was supported in part by the NSF
under grants PHY--9802564 and PHY--9722144.



\newpage

\end{document}